%% file: fan.tex
\documentclass[preprint2]{aastex}
\received{}
\accepted{}
\shorttitle{The Optical Nature of 62 X-ray Globular Clusters in M31}
\shortauthors{Fan et al.}

\begin{document}

\title{Identification of X-ray Point Sources and Study on the
Nature of 62 X-ray Globular Cluster Candidates in M31}

\author{
Zhou Fan\altaffilmark{1},
Jun Ma\altaffilmark{1},
Xu Zhou\altaffilmark{1},
Jiansheng Chen\altaffilmark{1},
Zhaoji Jiang\altaffilmark{1},
Zhenyu Wu\altaffilmark{1}
}
\altaffiltext{1}{National Astronomical Observatories,
Chinese Academy of Sciences, Beijing, 100012, P. R. China;
majun@bac.pku.edu.cn}

\begin{abstract}

This paper includes two parts. The first is to present the
spectral energy distributions (SEDs) of 49 globular cluster (GC)
X-ray sources in the BATC 13 intermediate-band filters from 3800
to 10000 \AA, and identify 8 unidentified X-ray sources in M31.
Using the X-ray data of $Einstein$ observation from 1979 to 1980,
$ROSAT$ HRI observation in 1990, $Chandra$ HRC and ACIS-I
observations from 1999 to 2001, and the BATC optical survey from
1995 to 1999, we find 49 GC X-ray sources and 8 new unidentified
X-ray sources in the BATC M31 field. By analyzing SEDs and FWHMs,
4 of the 8 X-ray sources may be GC candidates. The second is to
present some statistical relationships about 62 GC X-ray sources,
of which 58 are already known, and 4 are identified in this paper.
The distribution of M31 GC X-ray sources' V mags is bimodal, with
peaks at $ m_{\rm v} = 15.65$ and $m_{\rm v}= 17.89$, which is
different from the distribution of GC candidates. The distribution
of B-V color shows that, the GC X-ray sources seem to be
associated preferentially with the redder GCs, in agreement with
the previous results. Kolmogorov-Smirnov (KS) test shows that the
maximum value of the absolute difference of B-V distributions of
GC X-ray sources and GCs is $D_{max} = 0.181$, and the probability
$P = 0.068$ which means we can reject the hypothesis that the two
distributions are the same at the $90.0\%$ confidence level. In
the end, we study the correlation between X-ray luminosity (0.3-10
keV) and the optical luminosity (in V band) of the GC X-ray
sources in M31, and find that there exits a weak relationship with
the linear correlation coefficient $r = 0.36$ at the confidence
level of $98.0\%$.
\end{abstract}


\section{Introduction}

In our Local Group, M31 is the nearest (780 kpc) \citep{
stanekgarn98, mac01} and largest spiral galaxy with hundreds of
X-ray point sources \citep{su97, su01} in it. At the same time,
M31 possesses much more globular clusters (GCs) than the Milky Way
does, with the 435 GCs and GC candidates \citep{bh00}. As we know,
GC system can provide an important clue in our understanding of
the evolution and history of galaxies. It is also clear that GCs
are ideal laboratories for studying populations and the evolution
of dense stellar systems \citep{trpr04}. X-ray surveys reveal that
a number of bright X-ray sources associated with the Milky Way GCs
are identified as low-mass X-ray binaries (LMXBs) \citep{her83}.
The ratio of LMXBs to stellar mass is two orders of magnitude
higher for GCs than for the rest of our Galaxies
\citep{trpr04,liu01}. These LMXBs can be formed by capture from
the remnants of massive single stars that exploded with sufficient
isotropy to remain bound in GCs \citep{cla75}, or via tidally
dissipative two body encounters in the dense cores of GCs
\citep{fapr75}. However, the number of known Galactic GCs hosting
bright X-ray sources is quite small, for example, there are only
14 X-ray GCs with luminosities above $10^{36}\rm{erg~s^{-1}}$ in
\citet{liu01}. The X-ray observations of M31 have been carried out
many times and hundreds of X-ray point sources have been
discovered \citep{vans79,co90,trfab91,pri93,tr99,trfab91,su97,
su01, os01,ka02,kong02,wi04}. With these observations, many
optical counterparts of X-ray point sources in M31 have been
detected. Of all these counterparts, GCs may be important compared
with  GCs in our Milky Way. For example, in M31 about $1/3$ of the
GC X-ray sources have luminosity $L_{X} \geq 10^{37}\rm{erg~s^{-1}}$
(0.5-7.0 keV) as compared to about $1/12$ within our
Galaxy \citep{dist01}, and more than $30\%$ of the GC X-ray
sources in M31 have luminosity $L_{X}\geq 5 \times 10^{37}\rm{erg
~s^{-1}}$ as compared to only one in our Galaxy \citep{liu01,
sido01}. \citet{dist03} found that the optically bright X-ray GCs
house only the brightest X-ray sources, and that X-ray sources are
most likely to be found in red GCs.

The study on the correlation between optical and X-ray properties
for GC X-ray sources is also important. At visual wavelengths,
giants dominate the flux of GCs, and low-mass main-sequence stars
will show up with deeper exposures, and X-ray observations detect
compact objects \citep{verb95}. \citet{bell95} found that both
Galactic and M31 GCs housing bright LMXBs are both denser and more
metal-rich. Then \citet{trpr04} confirmed the conclusions of
\citet{bell95}.

M31 was optically observed as part of galaxy calibration program
of the Beijing-Arizona-Taiwan-Connecticut (BATC) Multicolor Sky
Survey \citep[e.g.,][]{fan96,zheng99}, which has a custom-designed
set of 15 intermediate-band filters to do spectrophotometry for
preselected 1 deg $^{2}$ regions of the northern sky.

The outline of this paper is as follows. The X-ray data used are
introduced in \S~2. Details of optical observations and data
reduction are given in \S~3. Optical counterparts of X-ray point
sources are identified in \S~4. The nature of 62 GC X-ray sources
is shown in \S~5. Finally, the summary is presented in \S~6.

\section{The Data of X-ray Point Sources in M31}

Globular clusters of X-ray point sources in this paper are mainly
from \citet{trpr04}, who presented the results of M31 GC X-ray
point sources survey based on the data of $XMM-Newton$ and
$Chandra$ observations covering $\sim 6100~{\rm arcmin^{2}}$ of
M31, and detected 43 GC X-ray point sources. At the same time, in
order to enlarge the GC X-ray point sources, we also refer to \citet
{trfab91,pri93,tr99,su01,ka02,kong02,wi04}. \citet{trfab91}
reported the result of entire set of $Einstein$ imaging
observations of M31 and detected 108 individual X-ray sources
(0.2-4.0 keV). \citet{pri93} gave us a report of a 48,000 s
observation of the central $\sim 34 \arcmin$ of M31 with $ROSAT$
HRI in 1990 and detected 86 X-ray sources with luminosities above
$\sim 1.4 \times 10^{36}\rm{erg~s^{-1}}$, of which 18 sources were
identified to be GCs. \citet{tr99} presented the spectral study of
the X-ray emitting stellar sources in M31, and found that GC
sources have spectral characteristics consistent with those of the
Milky Way objects. \citet{su01} analyzed the second $ROSAT$ survey
of M31 within the $\sim10.7 \rm{deg^{2}}$ field of view and
detected 396 individual X-ray point sources. Combined with the
first survey, they presented 560 X-ray point sources, 55 of which
are identified to be the foreground stars, 33 GCs, 16 SNRs and 10
radio sources and galaxies. Using the data on a deep observation
of the core of M31 made with the $Chandra$ High Resolution Camera
(HRC) \citep{murr97}, \citet{ka02} detected 143 X-ray point
sources, 20 of which were identified to be GC candidates. By
combining eight $Chandra$ ACIS-I observations taken between 1999
and 2001, \citet{kong02} detected 204 X-ray sources within the
central $\sim 17\arcmin\times17\arcmin$ region, with a detection
limit of $\sim 2.0 \times 10^{35}\rm{erg~s^{-1}}$. Of these X-ray
sources, \citet{kong02} identified 45 counterparts: 22 GCs, 9 PNe,
2 SNRs, 9 SSSs and 3 stars. \citet{wi04} discovered 166 X-ray
point sources, including 28 GCs, 17 Stars, 1 BL, 6 PNe and 3 SNRs,
from the 17 epochs of $Chandra$ HRC snapshot images, each covering
most of the M31 disk. By comparing these papers, we find that
there are 54 different GC X-ray point sources. Besides, we will
identify the optical counterparts of X-ray point sources using the
X-ray data of M31 from \citet{trfab91, pri93, tr99, ka02, kong02,
wi04} who presented the positions of X-ray point sources, and
using the BATC optical observation of M31.

\section{The BATC Optical Observation of M31}

The optical observations of M31 were carried out by BATC
Multicolor Sky Survey System, which uses a 60/90 cm f/3 Schmidt
telescope at Xinglong Station of the National Astronomical
Observatories of Chinese Academy Sciences (NAOC), where the seeing
is about $\sim2\arcsec$. A Ford Aerospace $\rm{2k} \times \rm{2k}$
large area CCD cameral is mounted at the Schmidt telescope focus,
giving a field of $58\arcmin\times58\arcmin$ with a pixel size of
$\sim1.67\arcsec$. This system includes 15 intermediate-band
filters, covering a range of wavelength from 3,000 to 10,000 \AA.
In our work, only 13 intermediate-band filters (BATC03 - BATC15,
$\sim 4,000 - 10,000$ \AA) are used. The optical observations of
M31 were carried out from Nov. 15th, 1995 to Dec. 16th, 1999 and
the total exposure lasted about 37 hours.

The calibrations of the images are made using observations of four
$F$ sub-dwarfs, HD~19445, HD~84937, BD~${+26^{\circ}2606}$, and
BD~${+17^{\circ}4708}$, all taken from \citet{ok83}. Hence, our
magnitudes are defined in a way similar to the spectro-photometric
AB magnitude system that is the Oke \& Gunn $\tilde{f_{\nu}}$
monochromatic system. BATC magnitudes are defined on the AB
magnitude system as

\begin{equation}
m_{\rm batc}=-2.5{\rm log}\tilde{F_{\nu}}-48.60,
\end{equation}

\noindent where $\tilde{F_{\nu}}$ is the appropriately averaged
monochromatic flux in unit of $\rm {erg~s^{-1} cm^{-2} Hz^{-1}}$
at the effective wavelength of the specific passband. In the BATC
system \citep{yan00}, $\tilde{F_{\nu}}$ is defined as

\begin{equation}
\tilde{F_{\nu}}=\frac{\int{d} ({\rm log}\nu)f_{\nu}r_{\nu}}
{\int{d} ({\rm log}\nu)r_{\nu}},
\end{equation}

\noindent which links the magnitude to the number of photons
detected by the CCD rather than to the input flux of Vega
\citep{fuku96}. In equation (2), $r_{\nu}$ is the system's
response, $f_{\nu}$ is the SEDs of the source.

Data reduction, beginning with the usual CCD processing steps of
bias subtraction and flat-fielding with dome flats, was performed
with an automatic data reduction software, PIPELINE I, developed
for the BATC Multicolor Sky Survey \citep{fan96, zheng99}.  The
dome flat-field images were taken by using a diffuser plate in
front of the correcting plate of the Schmidt telescope, a flat
field technique which has been verified with the photometry we
have done on other galaxies and field of view
\citep[e.g.,][]{fan96,zheng99,wu02,yan00,zhou01,zhou04}.
Spectrophotometric calibration of the M31 images using the
Oke-Gunn standard stars is done during photometric nights
\citep[see details from][]{yan00, zhou01}.

Using the images of the standard stars observed on photometric
nights, we derived iteratively the atmospheric extinction curves
and the variation of these extinction coefficients with time
\citep[cf.][]{zhou01}. The extinction coefficients at any given
time in a night $[K+ \Delta K (UT)]$ and the zero points of the
instrumental magnitudes ($C$) are obtained by

\begin{equation}
m_{\rm batc}=m_{\rm inst}+[K+\Delta K(UT)]X+C,
\end{equation}

\noindent where $X$ is the air mass. The instrumental magnitudes
($m_{\rm inst}$) of the selected bright, isolated and unsaturated
stars on the M31 field images of the same photometric nights can
be readily transformed to the BATC AB magnitude system ($m_{\rm
batc}$). The calibrated magnitudes of these stars are obtained on
the photometric nights, which are then used as secondary standards
to uniformly combined images from calibrated nights to those taken
during non-photometric weather. Table~1 lists the parameters of
the BATC filters and the statistics of observations. Column 6 of
Table~1 gives the scatter, in magnitudes, for the photometric
observations of the four primary standard stars in each filter.

\section{Identification of 4 New GC X-ray Sources}

\subsection{Finding Optical Counterparts of the GC X-ray Sources}

\citet{trfab91} presented the positions of 108 $Einstein$ X-ray
sources; \citet{pri93} listed the positions of 86 $ROSAT$ HRI
X-ray sources; \citet{kong02}, \citet{ka02} and \citet{wi04}
presented the positions of 204, 143 and 166 X-ray point sources
using $Chandra$ ACIS-I or HRC X-ray data, respectively. By
comparing the positions of these X-ray point sources with the
positions of objects in the BATC M31 field within a radius of
$(2^2+2^2)^{1/2}$ pixels, 17 X-ray point sources have optical
counterparts in the BATC CCD images except for the known optical
counterparts in these papers. We refer to the Simbad database, and
find that 9 X-ray sources have optical counterparts within a
radius of $5\arcsec$: CXOM31 J004216.0+411552, CXOM31
J004247.8+411052, CXOM31 J004301.7+411052, and RX J004138.3+410106
are identified as stars by \citet{berk88}; CXOM31 J004231.2+412008
is a GC candidate from \citet{gall04}; CXOM31 J004309.7+411901 and
CXOM31 J004304.2+411601 are also GC candidates from
\citet{wirth85}; [GMP2000b]J004240.7+405117.7 is a planetary
nebular from \citet{ford78}; 2E 0041.3+4114 is a GC from
\citet{bh00}. So, only 8 X-ray sources are needed to be
identified, and every X-ray source has only one optical
counterpart. Figure~1 shows these 8 optical counterparts, the
center of circles is the position of the X-ray sources, and the
radius is $(2^2+2^2)^{1/2}$ pixels.

\begin{figure}[htbp]
\figurenum{1}
\vspace{3.5cm} \caption{The
finding charts for 8 X-ray sources in M31. The centers of circles
are the positions of X-ray sources.}
\label{fig:one}
\end{figure}

\subsection{SEDs and FWHMs of GC X-ray Sources}

The BATC multicolor photometry can provide SED of an object. By
analyzing the SEDs and FWHM of an object in the M31 field, we can
determine whether an object is a GC candidate. In general, the
mean FWHM value of a M31 GC is wider than that of a star in the
Milky Way, and the SED of a GC do not vary steeply from one BATC
filter band to another and the fluxes are greater in longer
wavelength. We first present the SEDs and mean FWHMs of the known
GC X-ray point sources. Then, by comparing the SEDs and mean FWHMs
between the known GC X-ray point sources and the optical
counterparts of X-ray sources in the image of a BATC filter band,
we identify whether an optical counterpart of X-ray sources is a
GC candidate or not.

Using the positions of the GC X-ray sources from \cite{trfab91,
pri93, tr99, trpr04, kong02, wi04, su01} and referring the Simbad
database, we found 49 individual GC X-ray sources in the BATC M31
field. For each GC X-ray point source, the PHOT routine in DAOPHOT
\citep{stet87} is used to obtain magnitudes. To avoid
contamination from nearby objects, we adopt a small aperture of
$10.2\arcsec$ corresponding to a diameter of 6 pixels in the Ford
CCD. The large-aperture measures on the uncrowded bright stars
were used to determine the aperture corrections, i.e., the
magnitude difference between the small-aperture magnitude and the
``total'' or seeing-independent magnitude for the stars on each
frame. SEDs for 49 GC X-ray point sources in 13 BATC filters are
obtained from these measurements. These data are given in Table~2.
Column 2 to column 14 give the magnitudes of the 13 BATC passbands
observed. The second line for each GC gives the $1-\sigma$ errors
in magnitudes for the corresponding passband. The errors for each
filter are given by DAOPHOT. Figure~2 plots the SEDs for these 49
GC X-ray sources. For convenience, in Figure~2 the flux ratios
relative to filter BATC08 ($\lambda=6075${\AA}) are used except
for source Bo D91,MIT236 and source MIT311. For these two GCs, the
flux ratios relative to BATC10 ($\lambda=7010${\AA}) are used since
they are very red.

\begin{figure}[htbp]
\figurenum{2} \epsscale{0.80} \hspace{-0.2cm}
\rotatebox{-90}{\plotone{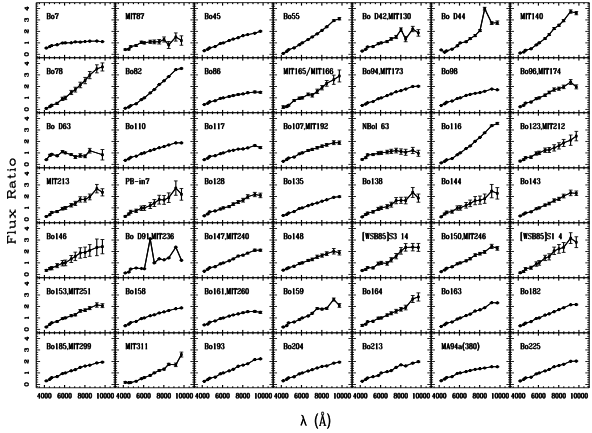}} \vspace{-0.3cm} \caption{SEDs of
49 known GC X-ray point sources in M31.} \label{fig:two}
\end{figure}

In Table~3, we list the mean FWHM values of column and line for
the 49 GC X-ray point sources in the BATC08 filter band. The mean
FWHM values for MIT213 and PB-in7 cannot be fitted because of low
signal-to-noise ratio or strong nearby background. We also list
the mean FWHM values of column and line for 8 Milky Way foreground
stars around the GC X-ray point sources. The errors of FWHMs are
the rms of FWHMs from 8 Milky Way foreground stars. From Table~3,
we can see that most of the GCs have larger FWHM values than
those of nearby foreground stars.

\subsection{SEDs and FWHMs of 8 New Optical Counterparts of X-ray
Point Sources}

For each optical counterpart of X-ray point sources, the PHOT
routine in DAOPHOT \citep{stet87} is used to obtain magnitudes
(see details from 4.2). SEDs for these optical counterparts in 13
BATC filters are obtained from these measurements, and listed in
Table~4. The magnitudes in 13 BATC passbands are from columns 2 to 14.
The second line lists the $1-\sigma$ errors of magnitude for the
corresponding passband, which are obtained from DAOPHOT. Figure~3
plots the SEDs for these 8 optical counterparts in 13 BATC
filters. For convenience, we also calculated the flux ratios
relative to BATC08 filter ($\lambda=6075$ {\AA}) as in Figure~2
except for CXOM31 J004237.9+410526. For this optical counterpart,
the flux ratios relative to BATC10 filter ($\lambda=7010${\AA})
are used since the source is very red.

\begin{figure}[htbp]
\figurenum{3} \epsscale{0.8} \hspace{-0.5cm}
\rotatebox{-90}{\plotone{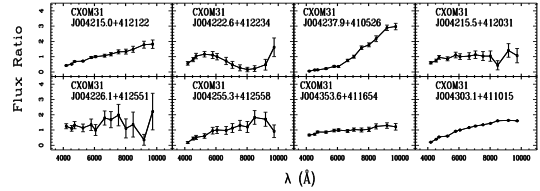}} \vspace{-2.0cm} \caption{SEDs of
the 8 unidentified optical counterparts of X-ray point sources.}
\label{fig:three}
\end{figure}

Table~5 lists the mean FWHM values of column and line for the
optical counterparts in BATC08 filter band. Column 1 lists the
name of X-ray sources. Columns 2 and 3 list R.A. and Dec. of the
sources, respectively, and column 4 is the mean FWHM values (for
column and line) of the optical counterparts of the X-ray sources.
Column 5 lists the mean FWHM values of column and line for 8 Milky Way
foreground stars around every X-ray point source. The errors of
FWHMs are the rms of FWHMs from 8 Milky Way foreground stars.
Column 10 lists the radial offsets between the X-ray sources and
their optical counterparts. For CXOM31 J004222.6+412234 and CXOM31
J004255.3+412558, which are two very faint objects, and the
background around them is very strong, it is uncertain to estimate
their FWHMs, so we did not estimate these two objects' FWHMs, and
did not identify them. So, there are 6 X-ray sources left to be
identified. We use the 49 known GC X-ray sources as template SEDs,
and compare the SED of each optical counterpart with all the 49
template SEDs by the $\chi^2$ method and choose the one with the
minimum value of $\chi^2$,

\begin{equation}
\chi^2=\sum_{i=3}^{15}{\frac{[C_{\lambda_i}^{\rm
new}(m)-C_{\lambda_i}^{\rm temp}(n)]^2}
{\sigma_{i}^{2}(m)+\sigma_{i}^{2}(n)}},
\end{equation}

\noindent where $C_{\lambda_i}^{\rm temp}(n)$ represents the flux
ratio in the $i$th filter (relative to filter BATC08
($\lambda=6075${\AA})) of the $n$th template GC X-ray source and
${\sigma_i}^{\rm temp}(n)$ is the corresponding observational flux
uncertainty in the $i$th filter of the $n$th template GC X-ray
source obtained from the magnitude uncertainty.
$C_{\lambda_i}^{\rm new}(m)$ and ${\sigma_i}^{\rm temp}(m)$ have
the similar meaning but for the $m$th new optical counterparts of
the X-ray source that need to be identified. Besides, we have
calculated the probability $Q$. The minimum values of $\chi^2$
($\chi^2_{min}$) and the corresponding $Q$ are listed in columns 7
and 8 of Table~5, respectively. We identify a new optical
counterpart of X-ray point sources to be a GC candidate as below.
First, when $Q$ is smaller than 0.1, we reject the hypothesis that
the candidate SED and the template are consistent as
\citet{lampton76} argued. Second, the mean FWHM value of a X-ray
GC candidate should be no less than those of the nearby foreground
stars. The optical counterparts identified as GC candidates are
indicated in column 9 and the names of the template GC X-ray
source with the minimum value of $\chi^2$ are listed in column 6.

\section{The Properties of 62 GC Candidates of X-ray Sources}

In this section, we will study the properties of 62 GC X-ray sources.

\subsection{The Sample of 62 GC Candidates of X-ray Sources}

There are 58 known GC candidates of X-ray sources found in the
literatures: 43 of which are from \citet{trpr04}; 7 from
\citet{su01}; 2 from \citet{wi04}; 1 from \citet{kong02}; 1 from
\citet{kong02} and \citet{gall04}; 1 from \citet{ka02} and
\citet{wirth85}; 1 from \citet{wi04} and \citet{wirth85}; 1 from
\citet{pri93}; 1 from \citet{trfab91} and \citet{bh00}. In order
to enlarge the sample of GC candidates of X-ray sources, we also
add 4 GC candidates identified from this paper. Therefore, there
are 62 GC candidates of X-ray sources in all. Figure~4 shows the
space distribution of them in the BATC M31 field (the circles
represent the known GC X-ray sources, while the boxes are
represent the new identified GC X-ray candidates). And there are 9
sources are out of the BATC M31 field or too dim to be detected,
so only 53 are shown in Figure~4.

\begin{figure}[htbp]
\figurenum{4}
\vspace{3.5cm} 
\caption{ The
position distributions of the GC X-ray sources in the BATC M31
field. The circles represent the known GC X-ray sources, while the
boxes represent the new identified GC X-ray candidates}
\label{fig:four}
\end{figure}

In Table~6, we list some parameters of the 58 known GC candidates
of X-ray sources from
\citet{trfab91,pri93,su01,ka02,kong02,wi04,trpr04}, and the 4 new
ones identified in this paper. Column 1 is the number; column 2
lists the optical names and column 3 are the names of X-ray
sources; column 4 lists the X-ray luminosities, of which the unit
is $\rm 10^{35}erg~s^{-1}$. The X-ray sources from \citet{trpr04}
are observed in the 0.3 - 10 keV energy band, from \citet{su01} in
the 0.1 - 2 keV, from \citet{kong02} and \citet{wi04} in the 0.3 -
7 keV, from \citet{ka02} in the 0.1 - 10 keV, and from
\citet{trfab91, pri93} in 0.2 - 4.0 keV.

\subsection{B,V Magnitude and B-V Color of 58 GC Candidates of X-ray Sources}

Using the Landolt standards, \citet{zhou03} presented the
relationships between the BATC intermediate-band system and UBVRI
broadband system from the catalogs of \citet{lan83,lan92} and
\citet{gala00}. We show the coefficients of two relationships in
equations (5) and (6):
\begin{equation}
m_{\rm B} = m_{04} + 0.2201 (m_{03} -m_{05}) + 0.1278\pm 0.076,
\end{equation}
\begin{equation}
m_{\rm V} = m_{07} + 0.3292 (m_{06} -m_{08}) + 0.0476\pm 0.027.
\end{equation}
The uncertainties in B (BATC) and V (BATC) are calculated as
following, $\sigma_{\rm B}$ = $(\sigma_{04}^{2} + 0.2201^{2}
(\sigma_{03}^{2} + \sigma_{05}^{2}))^{1/2}$, and $\sigma_{\rm V}$
= $(\sigma_{07}^{2} + 0.3292^{2}(\sigma_{06}^{2} +
\sigma_{08} ^{2}))^{1/2}$, to reflect the errors in the three filter
measurements. For the colors, we add the errors in quadrature,
i.e. $\sigma_{\rm {B-V}}$ = $(\sigma_{\rm B}^{2} + \sigma_{\rm
V}^{2})^{1/2}$.

Using the equations above, we transformed the magnitudes of 53 GC
candidates of X-ray sources in BATC03, BATC04 and BATC05 bands to
ones in B band, and in BATC06, BATC07 and BATC08 bands to ones in
V band. For the other 9 sources out of the BATC M31 field or
cannot be detected in the BATC M31 field, we can find V mags and
B-V colors of the 5 ones from \cite{batt87}. In all, there are 58
GC X-ray sources with V mags and B-V colors.

The fact that the GCs hosting X-ray sources tend to be optically
brighter than the rest of the M31 GCs, has been found by
\citet{dist02} and \citet{trpr04}. \citet{sarazin03} also found a
strong tendency for the X-ray sources to be associated with the
optically more luminous GCs in 4 early-type galaxies. With the
more sample GC X-ray sources, Figure~5 plots the distribution of V
mags for the GC candidates of X-ray sources along with the
distribution for the whole sample of the A and B class GC
candidates from \cite{batt87}. For seeing clearly, we use the
normalized number, i.e., in each bin the number of objects is
divided by the number of all the objects. From this figure, we can
see that these two distributions are not the same. Majority of GC
X-ray sources are brighter than 17 mag, which is the peak of
distribution of the GC candidates. This conclusion is in agreement
with one found by \citet{dist02} and \citet{trpr04}. At the same
time, the V mag distribution of the GC X-ray sources can be
clearly seen to be bimodal, i.e., besides the major brighter GCs
hosting X-ray sources that confirm the conclusions of
\citet{dist02} and \citet{trpr04}, there are also some GC X-ray
sources to be less luminous. As for the fact that the GCs hosting
X-ray sources tend to be optically brighter than the rest of the
M31 GCs, seems that it results primarily from the larger number of
stars in optically luminous GCs \citep{sarazin03}. As for the
result that there are some less luminous GC X-ray sources, it is
interesting but still need to be confirmed with a larger sample.

To make quantitative statements about the bimodality of V mags for
GC X-ray sources, the KMM test \citep{abz94} is applied to the
data. This test uses a maximum likelihood method to estimate the
probability that the data distribution is better modelled as a sum
of two Gaussians rather than a single Gaussian. Here we use a
homoscedastic test, i.e., the two Gaussians are assumed to have
the same dispersion. The $m_{\rm v}$ of the two peaks, the
$P$-value, and the numbers of GC X-ray sources assigned to each
peak by the KMM test are $m_{\rm v}\approx 15.65$ and 17.89,
0.094, 44 and 14, respectively. The $P$-value is in fact the
probability that the data are drawn from a single Gaussian
distribution. The KMM test suggests that we can consider the
distribution is bimodal at the $90.6\%$ confidence level. At the
same time, we use Kolmogorov-Smirnov (KS) test to demonstrate
whether the two distributions in Figure 5 are the same. We
determined a value of $D_{max}=0.354$ for these two samples with
58 and 333 points respectively. The probability of obtaining a
value of $D_{max}=0.354$ for 58 and 333 points is less than
$10^{-5}$. We can reject the hypothesis that the two distributions
are the same at $100\%$ confidence level. In addition, the KS test
determined a value of $D_{max}=0.566$ for the distribution of GC
X-ray sources between in \cite{trpr04} and in this paper. The
probability of obtaining a value of $D_{max}=0.566$ for 6 and 15
points is $7.89\%$. We can reject the hypothesis that the two
distributions are the same at $90\%$ confidence level. In the
latter KS test, we used the numbers of X-ray GCs in various bins
since \cite{trpr04} did not list the values of V mags in Table 3,
and the numbers are normalized, i.e., in each bin the number of
objects is divided by the number of all the objects.

\begin{figure}[htbp]
\figurenum{5} \epsscale{0.9} \hspace{-1.0cm}
\rotatebox{-90}{\plotone{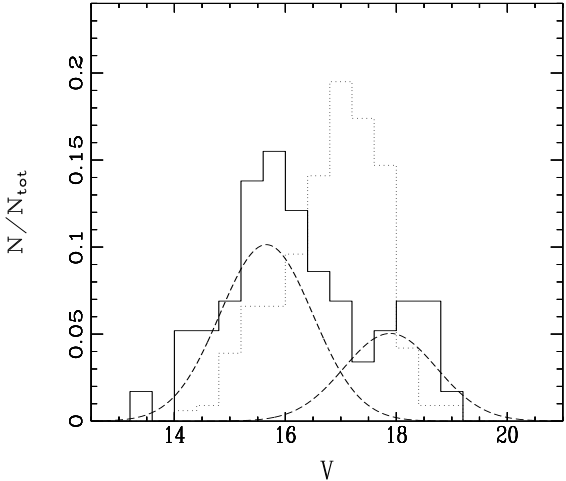}} \vspace{-0.5cm} \caption{The V
magnitude distributions of GC X-ray sources and all the GC
candidates in M31. The results for the GC X-ray sources are shown
with solid-line histogram. The distribution for the GC candidates
from \citet{batt87} is shown with dotted-line histogram. The
dashed lines show the bimodal fitting of V magnitude distribution of
GC X-ray sources, according to the KMM test.}\label{fig:five}
\end{figure}

In Figure~6, we plot the B-V color and V magnitude diagram
for the GC candidates of X-ray sources. We do not see any
evident difference with Figure~5 of \cite{batt87}.

\begin{figure}[htbp]
\figurenum{6} \epsscale{0.9} \hspace{-0.8cm}
\rotatebox{-90}{\plotone{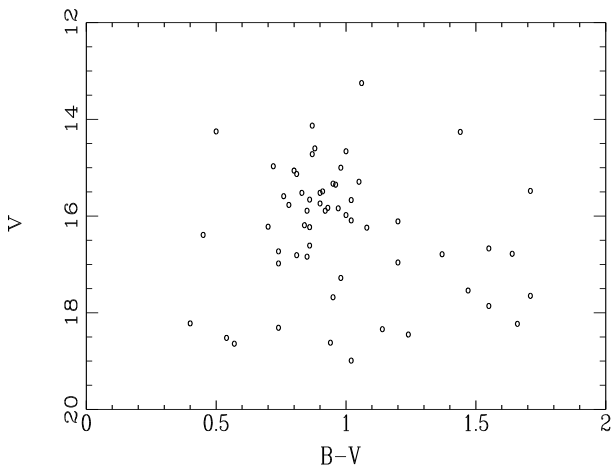}} \vspace{-0.5cm} \caption{The
B-V color and V magnitude diagram for the GC X-ray sources
in M31.} \label{fig:six}
\end{figure}

Using a sample of four early-type galaxies with X-ray sources,
\citet{sarazin03} found a tendency that the X-ray sources to be
found preferentially in redder GCs, which seems to indicate that
the evolution of X-ray binaries in GCs is affected by either the
metallicity or the age of the GC \citep{sarazin03}. Figure~7 plots
the distribution of B-V color for the GC candidates of X-ray
sources and the A, B class GC candidates of \citet{batt87}. For
seeing clearly, we also use the normalized number as in Figure 5.
It presents that the GC X-ray sources seem to be associated
preferentially with the redder GCs, in agreement with the results
of \citet{sarazin03}. In order to determine whether GC X-ray
sources and GCs are drawn from the same distribution, we also
provide the KS test. We determined a value of  $D_{max}=0.181$ for
these two samples with 58 and 328 points respectively. The
probability of obtaining a value of $D_{max}=0.181$ for 58 and 328
points is $6.8\%$, which means we can reject the hypothesis
that the two distributions are the same at $90\%$ confidence level.

\begin{figure}[htbp]
\figurenum{7} \epsscale{0.9} \hspace{-1.0cm}
\rotatebox{-90}{\plotone{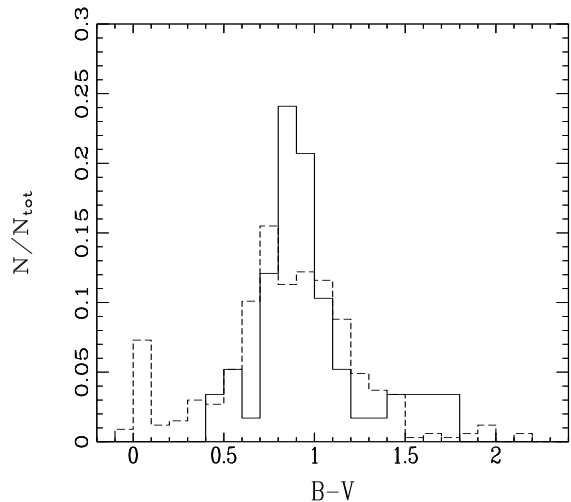}} \vspace{-0.5cm} \caption{The B-V
color distributions of the GC X-ray sources and the GC candidates
in M31. The results for the GC X-ray sources are shown with
solid-line histogram. The distribution for the GC candidates from
\citet{batt87} is shown with dashed-line histogram.}
\label{fig:seven}
\end{figure}

\subsection{Correlation between X-ray and Optical Luminosity
for the GC Candidates of X-ray Sources in M31}

Since the clusters of galaxies have the tight correlation between
the total optical luminosity and the X-ray luminosity $L_{\rm opt}
\propto L_{\rm X}^{0.45 \pm 0.03}$ \citep{pop04}, we wonder
whether the similar correlation exits between optical luminosity
and X-ray luminosity for GC X-ray sources. As the X-ray
luminosities of 62 sources are from several different observations
in different energy bands. Thus, we have to use the only 43 X-ray
GC candidates from \citet{trpr04} which were observed in the
0.3-10 keV energy band. A large number of X-ray GC sources have
X-ray luminosities in a wide range, so we use error bars in
Figure~8 to show the X-ray luminosity range as \citet{trpr04} did.
Besides, in Figure~8, 10\% uncertainty in X-ray luminosity for
some sources is accepted, which is translated directly from
uncertainty ($\pm 10\%$) in flux reported by \citet{trpr04}. There
are only two candidates, the V mags of which can not be obtained
(see 5.2 for details). So, Figure~8 shows the correlation between
X-ray luminosity and the V magnitude for 41 GC X-ray sources. For
presenting the quantitative analysis, we do the ordinary
least-squares fit. The fit formula is below,
\begin{equation}
log(L_{\rm X}/10^{35}) = a~m_{\rm v} + b,
\end{equation}
and the fit results are $a = -0.26\pm0.11, b= 6.17\pm1.76$, and the
linear correlation coefficient $r=0.36$ at the confidence level of
$98.0\%$. If our sample does not include the source No.61 in Table~6,
the fit results would be much better: $a = -0.31\pm0.10,
b= 6.96\pm1.64$ and the linear correlation coefficient $r= 0.45$ at
the confidence level of $99.5\%$.

\begin{figure}[htbp]
\figurenum{8} \epsscale{0.9} \hspace{-0.9cm}
\rotatebox{-90}{\plotone{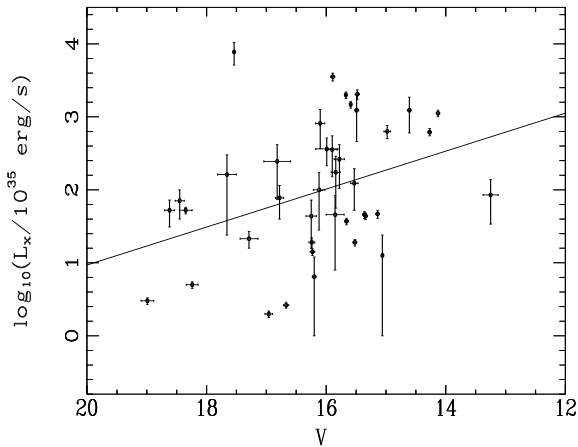}} \vspace{-0.5cm} \caption{The
correlation between X-ray luminosities and V mags for the GC X-ray
sources in M31.}\label{fig:eight}
\end{figure}

\section{Summary}

In this paper, we study the properties of GC X-ray sources.
Firstly, we identify the optical counterparts of the X-ray point
sources in M31 using the results of the $Chandra$ ACIS-I and HRC
observations from 1999 to 2001, $Einstein$ observation from 1979
to 1980, and $ROSAT$ HRI observation in 1990, and the BATC optical
survey from 1995 to 1999. We identify 8 new optical counterparts
of X-ray point sources, and find that 4 of them probably be GC
candidates by analyzing both the SEDs and FWHMs. Secondly, we
collect 62 GC X-ray sources, of which 53 GC X-ray sources' SEDs
are from 4000 to 10000 {\AA}, the V mags and B-V colors are
obtained using the BATC optical survey from 1995 to 1999. Then we
do some statistical relationships about the GC X-ray sources. The
results show that the distributions of V mag of the GC X-ray
sources and GC candidates are different. Majority of GC X-ray
sources are brighter than 17 mag, which is the peak of
distribution of the GC candidates. At the same time, the V
distribution of the GC X-ray sources in M31 is bimodal, with peaks
at $m_{\rm v}$ = 15.65 and 17.89. The distributions of B-V color
of GC X-ray sources and GC candidates \cite{batt87} show that the
GC X-ray sources seem to be associated preferentially with the
redder GCs, in agreement with the previous results, and the KS
test gives that the maximum value of the absolute difference of
these two distributions $D_{max} = 0.181$ and $P = 0.068$.
Finally, we study the correlation of the X-ray luminosity (0.3-10
keV) and the optical luminosity (V mag), and find there exits a
weak relationship with the linear correlation coefficient $r =
0.36$ at the confidence level of $98.0\%$.

\section{Acknowledgments}

We would like to thank the anonymous referee for his/her
insightful comments and suggestions that improved this paper very
much. This work has been supported by the Chinese National Key
Basic Research Science Foundation (NKBRSF TG199075402) and by the
Chinese National Natural Science Foundation, No. 10473012.

\clearpage
\input{table1.tex}
\clearpage
\input{table2.tex}
\clearpage
\input{table3.tex}
\clearpage
\input{table4.tex}
\clearpage
\input{table5.tex}
\clearpage
\input{table6.tex}
\end{document}

%% file: table1.tex
\begin{deluxetable}{cccccc}
\tablenum{1}
\tabletypesize{\footnotesize}
\tablewidth{0pt}
\tablecaption{Parameters of the BATC Filters and Statistics of Observations for M31}
\tablehead{\colhead{No.}&\colhead{Name}&\colhead{cw(\AA)\tablenotemark{a}}&\colhead{Exp. (hr)}&\colhead{N.img\tablenotemark{b}}&\colhead{rms\tablenotemark{c}}}
\tablecolumns{6}
\startdata
 1  & BATC03 & 4210   & 01:00 & 03 & 0.015\\
 2  & BATC04 & 4546   & 05:30 & 17 & 0.009\\
 3  & BATC05 & 4872   & 03:30 & 11 & 0.015\\
 4  & BATC06 & 5250   & 02:20 & 12 & 0.006\\
 5  & BATC07 & 5785   & 02:15 & 07 & 0.003\\
 6  & BATC08 & 6075   & 01:40 & 05 & 0.003\\
 7  & BATC09 & 6710   & 00:45 & 03 & 0.003\\
 8  & BATC10 & 7010   & 03:00 & 12 & 0.008\\
 9  & BATC11 & 7530   & 02:00 & 06 & 0.004\\
10  & BATC12 & 8000   & 04:00 & 12 & 0.003\\
11  & BATC13 & 8510   & 01:30 & 05 & 0.004\\
12  & BATC14 & 9170   & 05:50 & 18 & 0.003\\
13  & BATC15 & 9720   & 04:00 & 12 & 0.009\\
\enddata
\tablenotetext{a}{Central wavelength for each BATC filter.}
\tablenotetext{b}{Image numbers for each BATC filter.}
\tablenotetext{c}{Calibration error, in magnitude, for each filter as obtained from the standard stars.}
\end{deluxetable}

%% file: table2.tex
\begin{deluxetable}{cccccccccccccc}
\tablenum{2}
\tabletypesize{\footnotesize}
\tablewidth{0pt} \tablecaption{The SEDs of 49 X-ray GC
Candidates in 13 BATC Filter Bands.}
\tablehead{\colhead{}&\colhead{03}&\colhead{04}&\colhead{05}&\colhead{06}&\colhead{07}&\colhead{08}&\colhead{09}&\colhead{10}&\colhead{11}&\colhead{12}&\colhead{13}&\colhead{14}&\colhead{15}\\
\colhead{}&\colhead{4210}&\colhead{4546}&\colhead{4872}&\colhead{5250}&\colhead{5785}&\colhead{6075}&\colhead{6710}&\colhead{7010}&\colhead{7530}&\colhead{8000}&\colhead{8510}&\colhead{9170}&\colhead{9720}\\
\cline{2-14}
\colhead{Name}& \multicolumn{13}{c}{Central wavelength for each BATC filter (\AA)}}
\tablecolumns{14}
\startdata
 Bo7             & 14.78& 14.55& 14.42& 14.33& 14.14& 14.13& 14.12& 14.04& 14.06& 14.01& 13.98& 13.97& 14.02\\
 \nodata         & 0.011& 0.004& 0.003& 0.003& 0.003& 0.002& 0.002& 0.003& 0.003& 0.003& 0.004& 0.008& 0.013\\
 MIT87           & 19.85& 19.77& 19.33& 19.18& 18.85& 18.90& 18.80& 18.81& 18.80& 18.62& 19.12& 18.45& 18.70\\
 \nodata         & 0.155& 0.174& 0.115& 0.123& 0.116& 0.115& 0.148& 0.175& 0.218& 0.201& 0.510& 0.292& 0.455\\
 Bo45            & 16.75& 16.37& 16.11& 15.91& 15.55& 15.49& 15.29& 15.18& 15.09& 14.96& 14.90& 14.83& 14.73\\
 \nodata         & 0.057& 0.016& 0.010& 0.008& 0.007& 0.006& 0.004& 0.007& 0.007& 0.006& 0.008& 0.016& 0.024\\
 Bo55            & 18.99& 17.70& 17.19& 16.85& 16.42& 16.22& 15.94& 15.77& 15.59& 15.42& 15.27& 15.05& 14.99\\
 \nodata         & 0.407& 0.068& 0.040& 0.030& 0.028& 0.023& 0.021& 0.022& 0.022& 0.019& 0.021& 0.027& 0.036\\
 Bo D42,MIT130   & 19.70& 19.24& 18.81& 18.77& 18.43& 18.34& 18.22& 18.02& 17.89& 17.51& 18.03& 17.48& 17.66\\
 \nodata         & 0.091& 0.073& 0.058& 0.083& 0.084& 0.080& 0.099& 0.096& 0.107& 0.083& 0.211& 0.121& 0.178\\
 Bo D44          & 18.83& 19.62& 19.09& 18.66& 18.12& 17.79& 17.69& 17.45& 17.26& 17.00& 16.29& 16.70& 16.69\\
 \nodata         & 0.068& 0.190& 0.138& 0.097& 0.077& 0.051& 0.060& 0.054& 0.054& 0.045& 0.034& 0.048& 0.063\\
 MIT140          & 18.61& 18.02& 17.38& 17.05& 16.47& 16.26& 15.91& 15.67& 15.38& 15.25& 15.09& 14.83& 14.87\\
 \nodata         & 0.316& 0.125& 0.081& 0.069& 0.057& 0.050& 0.044& 0.044& 0.041& 0.038& 0.037& 0.044& 0.046\\
 Bo78,MIT153     & 20.01& 18.88& 18.39& 17.99& 17.37& 17.28& 16.86& 16.69& 16.46& 16.28& 16.09& 15.91& 15.86\\
 \nodata         & 0.521& 0.295& 0.262& 0.214& 0.177& 0.178& 0.146& 0.141& 0.133& 0.129& 0.121& 0.113& 0.110\\
 Bo82,MIT159     & 17.58& 16.76& 16.20& 15.78& 15.17& 14.99& 14.60& 14.40& 14.15& 14.00& 13.85& 13.64& 13.61\\
 \nodata         & 0.119& 0.028& 0.015& 0.012& 0.008& 0.007& 0.004& 0.006& 0.005& 0.004& 0.005& 0.007& 0.011\\
 Bo86,MIT164     & 15.72& 15.46& 15.22& 15.09& 14.82& 14.77& 14.65& 14.57& 14.49& 14.45& 14.37& 14.32& 14.34\\
 \nodata         & 0.036& 0.037& 0.042& 0.040& 0.039& 0.039& 0.039& 0.043& 0.048& 0.045& 0.046& 0.063& 0.055\\
 MIT165/MIT166   & 19.82& 19.62& 19.18& 18.34& 18.09& 18.05& 17.76& 17.83& 17.56& 17.36& 17.16& 17.02& 16.89\\
 \nodata         & 0.897& 0.367& 0.212& 0.106& 0.123& 0.109& 0.103& 0.142& 0.141& 0.117& 0.127& 0.182& 0.214\\
 Bo94,MIT173     & 16.42& 16.09& 15.84& 15.67& 15.33& 15.25& 15.05& 14.93& 14.79& 14.72& 14.63& 14.50& 14.49\\
 \nodata         & 0.041& 0.015& 0.011& 0.009& 0.009& 0.007& 0.005& 0.008& 0.008& 0.006& 0.009& 0.014& 0.022\\
 Bo98            & 17.17& 16.76& 16.49& 16.32& 16.03& 15.97& 15.76& 15.68& 15.61& 15.56& 15.48& 15.35& 15.39\\
 \nodata         & 0.011& 0.007& 0.007& 0.007& 0.007& 0.006& 0.007& 0.007& 0.008& 0.007& 0.014& 0.011& 0.017\\
 Bo96,MIT174     & 17.43& 16.97& 16.47& 16.31& 15.90& 15.82& 15.59& 15.40& 15.23& 15.18& 15.09& 14.88& 15.09\\
 \nodata         & 0.148& 0.118& 0.113& 0.111& 0.098& 0.095& 0.094& 0.093& 0.096& 0.087& 0.089& 0.102& 0.104\\
 Bo D63          & 19.45& 18.75& 18.62& 18.80& 18.37& 18.48& 18.66& 18.96& 18.76& 18.86& 18.29&\nodata& 18.67\\
 \nodata         & 0.068& 0.148& 0.092& 0.108& 0.093& 0.086& 0.096& 0.245& 0.240& 0.212& 0.201&\nodata& 0.841\\
 Bo110           & 16.03& 15.68& 15.41& 15.25& 14.96& 14.88& 14.69& 14.59& 14.48& 14.40& 14.31& 14.20& 14.20\\
 \nodata         & 0.029& 0.011& 0.008& 0.007& 0.007& 0.006& 0.005& 0.007& 0.007& 0.007& 0.008& 0.012& 0.017\\
 Bo117           & 16.96& 16.67& 16.39& 16.30& 16.07& 15.99& 15.82& 15.77& 15.68& 15.67& 15.60& 15.45& 15.58\\
 \nodata         & 0.068& 0.030& 0.016& 0.016& 0.031& 0.016& 0.012& 0.017& 0.017& 0.017& 0.025& 0.032& 0.061\\
 Bo107,MIT192    & 16.98& 16.46& 16.18& 16.02& 15.62& 15.53& 15.32& 15.24& 15.13& 15.03& 14.95& 14.84& 14.84\\
 \nodata         & 0.097& 0.075& 0.082& 0.081& 0.073& 0.069& 0.072& 0.079& 0.086& 0.076& 0.079& 0.104& 0.090\\
 NBol 63         & 17.53& 17.19& 16.81& 16.77& 16.63& 16.60& 16.52& 16.46& 16.40& 16.48& 16.54& 16.39& 16.64\\
 \nodata         & 0.108& 0.117& 0.119& 0.125& 0.156& 0.162& 0.169& 0.183& 0.199& 0.254& 0.312& 0.275& 0.348\\
 Bo116           & 18.56& 17.76& 17.32& 17.05& 16.50& 16.31& 15.96& 15.80& 15.55& 15.38& 15.21& 14.99& 14.93\\
 \nodata         & 0.313& 0.054& 0.029& 0.024& 0.021& 0.015& 0.011& 0.015& 0.014& 0.011& 0.014& 0.019& 0.030\\
 Bo123,MIT212    & 18.37& 17.97& 17.61& 17.51& 17.07& 17.01& 16.78& 16.76& 16.62& 16.49& 16.33& 16.20& 16.03\\
 \nodata         & 0.278& 0.180& 0.191& 0.200& 0.179& 0.174& 0.178& 0.208& 0.230& 0.203& 0.198& 0.259& 0.198\\
 MIT213          & 14.49& 13.97& 13.53& 13.40& 13.06& 12.98& 12.75& 12.62& 12.38& 12.38& 12.25& 11.91& 12.05\\
 \nodata         & 0.141& 0.147& 0.151& 0.151& 0.149& 0.147& 0.147& 0.152& 0.155& 0.153& 0.152& 0.162& 0.155\\
 PB-in7          & 15.95& 15.31& 14.92& 14.80& 14.51& 14.48& 14.28& 14.10& 13.89& 13.91& 13.79& 13.39& 13.63\\
 \nodata         & 0.237& 0.221& 0.238& 0.240& 0.247& 0.256& 0.262& 0.264& 0.275& 0.277& 0.276& 0.282& 0.289\\
 Bo128           & 17.85& 17.38& 17.03& 16.91& 16.68& 16.57& 16.45& 16.26& 16.12& 16.01& 15.83& 15.73& 15.77\\
 \nodata         & 0.094& 0.089& 0.086& 0.087& 0.098& 0.095& 0.098& 0.095& 0.101& 0.112& 0.103& 0.100& 0.110\\
 Bo135           & 16.83& 16.56& 16.28& 16.06& 15.70& 15.64& 15.45& 15.34& 15.25& 15.14& 15.06& 14.93& 14.90\\
 \nodata         & 0.060& 0.020& 0.014& 0.012& 0.011& 0.010& 0.009& 0.012& 0.013& 0.011& 0.014& 0.019& 0.030\\
 Bo138           & 17.20& 16.46& 16.20& 15.99& 15.64& 15.53& 15.42& 15.26& 15.00& 14.98& 14.99& 14.60& 14.86\\
 \nodata         & 0.228& 0.180& 0.213& 0.194& 0.187& 0.176& 0.192& 0.198& 0.194& 0.188& 0.213& 0.221& 0.234\\
 Bo144           & 17.44& 16.60& 16.32& 16.21& 15.79& 15.78& 15.56& 15.42& 15.25& 15.22& 15.14& 14.80& 14.89\\
 \nodata         & 0.268& 0.198& 0.235& 0.244& 0.225& 0.233& 0.237& 0.252& 0.271& 0.260& 0.277& 0.294& 0.269\\
 Bo143           & 16.91& 16.46& 16.23& 16.06& 15.69& 15.60& 15.36& 15.25& 15.06& 14.97& 14.83& 14.69& 14.71\\
 \nodata         & 0.102& 0.076& 0.091& 0.090& 0.090& 0.087& 0.086& 0.098& 0.103& 0.092& 0.092& 0.121& 0.100\\
 Bo146           & 17.77& 17.38& 17.29& 16.93& 16.66& 16.63& 16.31& 16.14& 15.95& 15.92& 15.83& 15.70& 15.67\\
 \nodata         & 0.254& 0.252& 0.345& 0.287& 0.300& 0.311& 0.280& 0.293& 0.304& 0.295& 0.301& 0.400& 0.325\\
 Bo D91,MIT236   & 15.81& 15.25& 14.36& 14.17& 14.26& 14.31& 12.30& 13.53& 13.20& 13.27& 13.14& 12.59& 13.31\\
 \nodata         & 0.024& 0.009& 0.004& 0.004& 0.005& 0.005& 0.001& 0.004& 0.004& 0.004& 0.004& 0.005& 0.009\\
 Bo147,MIT240    & 16.64& 16.11& 15.82& 15.68& 15.33& 15.24& 15.06& 14.92& 14.73& 14.68& 14.57& 14.43& 14.43\\
 \nodata         & 0.075& 0.055& 0.058& 0.056& 0.052& 0.047& 0.046& 0.049& 0.049& 0.043& 0.043& 0.054& 0.047\\
 Bo148           & 16.62& 16.31& 16.13& 15.90& 15.60& 15.53& 15.32& 15.24& 15.07& 14.99& 14.89& 14.77& 14.84\\
 \nodata         & 0.078& 0.071& 0.090& 0.086& 0.088& 0.088& 0.087& 0.100& 0.110& 0.099& 0.104& 0.136& 0.122\\
 $\rm [WSB85]S3~14$   & 17.48& 17.21& 16.86& 16.91& 16.36& 16.29& 16.34& 16.03& 15.79& 15.51& 15.36& 15.35& 15.36\\
 \nodata         & 0.119& 0.152& 0.157& 0.178& 0.162& 0.163& 0.194& 0.172& 0.169& 0.160& 0.162& 0.161& 0.174\\
 Bo150,MIT246    & 17.53& 16.99& 16.63& 16.41& 16.04& 15.95& 15.77& 15.61& 15.39& 15.31& 15.22& 14.98& 15.06\\
 \nodata         & 0.161& 0.117& 0.113& 0.097& 0.087& 0.082& 0.078& 0.081& 0.079& 0.071& 0.071& 0.085& 0.082\\
 $\rm [WSB85]S1~4$     & 19.64& 18.99& 18.32& 18.10& 17.62& 17.51& 17.12& 17.05& 16.73& 16.64& 16.58& 16.26& 16.39 \\
 \nodata         & 0.523& 0.427& 0.323& 0.279& 0.256& 0.243& 0.193& 0.209& 0.189& 0.205& 0.220& 0.173& 0.202 \\
 Bo153,MIT251    & 17.55& 16.74& 16.44& 16.25& 15.91& 15.84& 15.65& 15.58& 15.32& 15.28& 15.17& 15.02& 15.05\\
 \nodata         & 0.143& 0.078& 0.087& 0.079& 0.076& 0.073& 0.077& 0.085& 0.082& 0.078& 0.079& 0.102& 0.092\\
 Bo158           & 15.61& 15.20& 14.91& 14.75& 14.42& 14.35& 14.16& 14.09& 13.96& 13.89& 13.81& 13.71& 13.67\\
 \nodata         & 0.019& 0.007& 0.004& 0.004& 0.004& 0.003& 0.002& 0.004& 0.004& 0.003& 0.005& 0.007& 0.010\\
 Bo161,MIT260    & 17.03& 16.85& 16.52& 16.35& 16.07& 16.01& 15.85& 15.76& 15.68& 15.60& 15.54& 15.52& 15.58\\
 \nodata         & 0.078& 0.039& 0.035& 0.032& 0.032& 0.030& 0.030& 0.034& 0.038& 0.034& 0.039& 0.056& 0.069\\
 Bo159           & 18.29& 17.79& 17.19& 17.06& 16.75& 16.58& 16.41& 16.21& 15.93& 15.96& 15.92& 15.54& 15.78\\
 \nodata         & 0.241& 0.097& 0.067& 0.064& 0.058& 0.045& 0.043& 0.045& 0.043& 0.042& 0.048& 0.057& 0.086\\
 Bo164           & 18.62& 18.35& 17.92& 17.85& 17.51& 17.48& 17.24& 17.11& 16.98& 16.87& 16.78& 16.42& 16.35\\
 \nodata         & 0.288& 0.130& 0.100& 0.104& 0.105& 0.098& 0.094& 0.108& 0.119& 0.106& 0.123& 0.133& 0.145\\
 Bo163           & 16.18& 15.66& 15.34& 15.16& 14.79& 14.68& 14.48& 14.37& 14.17& 14.10& 13.99& 13.76& 13.77\\
 \nodata         & 0.033& 0.012& 0.009& 0.007& 0.007& 0.006& 0.005& 0.007& 0.007& 0.006& 0.007& 0.008& 0.013\\
 Bo182           & 16.38& 16.04& 15.75& 15.53& 15.16& 15.10& 14.86& 14.77& 14.63& 14.51& 14.41& 14.27& 14.26\\
 \nodata         & 0.038& 0.013& 0.008& 0.007& 0.006& 0.006& 0.004& 0.007& 0.007& 0.005& 0.007& 0.012& 0.017\\
 Bo185,MIT299    & 16.53& 16.11& 15.81& 15.66& 15.30& 15.24& 15.05& 14.97& 14.81& 14.73& 14.68& 14.56& 14.52\\
 \nodata         & 0.040& 0.015& 0.011& 0.010& 0.010& 0.009& 0.009& 0.011& 0.012& 0.011& 0.012& 0.018& 0.024\\
 MIT311          & 19.13& 19.37& 19.23& 18.68& 18.01& 17.82& 17.52& 17.26& 17.05& 16.95& 16.65& 16.68& 16.22\\
 \nodata         & 0.494& 0.305& 0.248& 0.155& 0.106& 0.086& 0.063& 0.066& 0.061& 0.060& 0.057& 0.106& 0.095\\
 Bo193           & 16.47& 15.98& 15.70& 15.53& 15.12& 15.03& 14.85& 14.73& 14.53& 14.47& 14.39& 14.19& 14.16\\
 \nodata         & 0.043& 0.012& 0.009& 0.007& 0.007& 0.005& 0.004& 0.006& 0.006& 0.005& 0.008& 0.010& 0.016\\
 Bo204           & 16.63& 16.24& 15.97& 15.82& 15.46& 15.37& 15.20& 15.10& 14.98& 14.88& 14.85& 14.70& 14.65\\
 \nodata         & 0.049& 0.016& 0.011& 0.009& 0.009& 0.008& 0.007& 0.009& 0.011& 0.009& 0.012& 0.018& 0.026\\
 Bo213           & 17.92& 17.42& 17.13& 16.94& 16.79& 16.51& 16.40& 16.31& 16.08& 15.93& 15.99& 15.84& 15.77\\
 \nodata         & 0.027& 0.022& 0.020& 0.017& 0.023& 0.017& 0.019& 0.018& 0.019& 0.018& 0.026& 0.024& 0.028\\
 MA94a(380)      & 15.73& 15.31& 15.03& 14.91& 14.53& 14.48& 14.36& 14.27& 14.19& 14.14& 14.09& 14.01& 14.01\\
 \nodata         & 0.021& 0.008& 0.005& 0.005& 0.005& 0.004& 0.003& 0.005& 0.005& 0.004& 0.006& 0.009& 0.015\\
 Bo225           & 15.12& 14.71& 14.42& 14.32& 13.91& 13.81& 13.63& 13.52& 13.34& 13.27& 13.20& 13.05& 13.04\\
 \nodata         & 0.014& 0.005& 0.004& 0.003& 0.003& 0.002& 0.002& 0.003& 0.003& 0.002& 0.003& 0.005& 0.007\\
\enddata
\end{deluxetable}

%% file: table3.tex
\begin{deluxetable}{ccc}
\tablenum{3}
\tabletypesize{\footnotesize}
\tablewidth{0pt} \tablecaption{The FWHM of 49 X-ray
GC Candidates in BATC08 Filter Band}
\tablehead{\colhead{Optical Name}&\colhead{FWHMs (GCs)}&\colhead{FWHMs (stars)}\\
\colhead{}&\colhead{Pixel}&\colhead{Pixel}}
\tablecolumns{3}
\startdata
 Bo7                & $ 2.5\pm0.14$ & $ 2.4\pm0.14$ \\
 MIT87              & $ 4.3\pm0.45$ & $ 2.7\pm0.45$ \\
 Bo45               & $ 2.7\pm0.05$ & $ 2.2\pm0.05$ \\
 Bo55               & $ 2.8\pm0.23$ & $ 2.1\pm0.23$ \\
 Bo D42,MIT130      & $ 3.5\pm0.18$ & $ 2.6\pm0.18$ \\
 Bo D44             & $ 3.8\pm0.02$ & $ 2.6\pm0.02$ \\
 MIT140             & $ 2.6\pm0.21$ & $ 2.4\pm0.21$ \\
 Bo78,MIT153        & $ 2.5\pm0.21$ & $ 2.4\pm0.21$ \\
 Bo82,MIT159        & $ 3.1\pm0.13$ & $ 2.6\pm0.13$ \\
 Bo86,MIT164        & $ 2.6\pm0.24$ & $ 2.3\pm0.24$ \\
 MIT165/MIT166      & $ 3.4\pm0.12$ & $ 2.0\pm0.12$ \\
 Bo94,MIT173        & $ 3.3\pm0.09$ & $ 2.5\pm0.09$ \\
 Bo98               & $ 3.0\pm0.09$ & $ 2.5\pm0.09$ \\
 Bo96,MIT174        & $ 3.4\pm0.10$ & $ 2.0\pm0.10$ \\
 Bo D63             & $ 3.1\pm0.02$ & $ 2.9\pm0.02$ \\
 Bo110              & $ 2.8\pm0.06$ & $ 2.5\pm0.06$ \\
 Bo117              & $ 3.3\pm0.09$ & $ 2.5\pm0.09$ \\
 Bo107,MIT192       & $ 2.3\pm0.10$ & $ 2.0\pm0.10$ \\
 NBol 63            & $ 2.8\pm0.10$ & $ 2.0\pm0.10$ \\
 Bo116              & $ 2.5\pm0.02$ & $ 2.1\pm0.02$ \\
 Bo123,MIT212       & $ 2.9\pm0.04$ & $ 2.4\pm0.04$ \\
 MIT213             & \nodata       & \nodata       \\
 PB-in7             & \nodata       & \nodata       \\
 Bo128              & $ 2.2\pm0.04$ & $ 2.4\pm0.04$ \\
 Bo135              & $ 2.4\pm0.24$ & $ 2.2\pm0.24$ \\
 Bo138              & $ 3.4\pm0.12$ & $ 2.3\pm0.12$ \\
 Bo144              & $ 2.7\pm0.03$ & $ 2.3\pm0.03$ \\
 Bo143              & $ 2.7\pm0.12$ & $ 2.3\pm0.12$ \\
 Bo146              & $ 3.5\pm0.03$ & $ 2.3\pm0.03$ \\
 Bo D91,MIT236      & $ 3.2\pm0.24$ & $ 2.1\pm0.24$ \\
 Bo147,MIT240       & $ 2.8\pm0.06$ & $ 2.1\pm0.06$ \\
 Bo148              & $ 2.7\pm0.08$ & $ 2.3\pm0.08$ \\
 $\rm[WSB85]S3~14$       & $ 2.4\pm0.03$ & $ 2.3\pm0.03$ \\
 Bo150,MIT246       & $ 2.8\pm0.06$ & $ 2.1\pm0.06$ \\
 $\rm[WSB85]S1~4$        & $ 2.7\pm0.08$ & $ 2.3\pm0.08$ \\
 Bo153,MIT251       & $ 2.4\pm0.03$ & $ 2.3\pm0.03$ \\
 Bo158              & $ 2.8\pm0.02$ & $ 2.3\pm0.02$ \\
 Bo161,MIT260       & $ 2.6\pm0.02$ & $ 2.3\pm0.02$ \\
 Bo159              & $ 3.0\pm0.04$ & $ 2.1\pm0.04$ \\
 Bo164              & $ 2.5\pm0.03$ & $ 2.3\pm0.03$ \\
 Bo163              & $ 2.6\pm0.04$ & $ 2.1\pm0.04$ \\
 Bo182              & $ 3.0\pm0.02$ & $ 2.3\pm0.02$ \\
 Bo185,MIT299       & $ 2.5\pm0.03$ & $ 2.3\pm0.03$ \\
 MIT311             & $ 2.8\pm0.03$ & $ 2.2\pm0.03$ \\
 Bo193              & $ 2.8\pm0.34$ & $ 1.9\pm0.34$ \\
 Bo204              & $ 2.5\pm0.11$ & $ 2.2\pm0.11$ \\
 Bo213              & $ 2.4\pm0.05$ & $ 2.4\pm0.05$ \\
 MA94a(380)         & $ 2.3\pm0.20$ & $ 2.2\pm0.20$ \\
 Bo225              & $ 2.7\pm0.01$ & $ 2.3\pm0.01$ \\
\enddata
\end{deluxetable}

%% file: table4.tex
\begin{deluxetable}{cccccccccccccc}
\tablenum{4}
\tabletypesize{\footnotesize}
\rotate \tablewidth{0pt} \tablecaption{The SEDs of 8 X-Ray
Optical Counterparts in 13 BATC Filter Bands}
\tablehead{\colhead{}&\colhead{03}&\colhead{04}&\colhead{05}&\colhead{06}&\colhead{07}&\colhead{08}&\colhead{09}&\colhead{10}&\colhead{11}&\colhead{12}&\colhead{13}&\colhead{14}&\colhead{15}\\
\colhead{}&\colhead{4210}&\colhead{4546}&\colhead{4872}&\colhead{5250}&\colhead{5785}&\colhead{6075}&\colhead{6710}&\colhead{7010}&\colhead{7530}&\colhead{8000}&\colhead{8510}&\colhead{9170}&\colhead{9720}\\
\cline{2-14}
\colhead{Name}& \multicolumn{13}{c}{Central wavelength for each BATC filter (\AA)}}
\tablecolumns{14}
\startdata
CXOM31 J004215.0+412122 & 19.02 & 18.80 & 18.49 & 18.45 & 18.14 & 18.08 & 18.01 & 17.92 & 17.78 & 17.77 & 17.65 & 17.45 & 17.43 \\
 \nodata    & 0.066 & 0.060 & 0.056 & 0.068 & 0.078 & 0.072 & 0.099 & 0.105 & 0.115 & 0.137 & 0.155 & 0.147 & 0.178 \\
CXOM31 J004222.6+412234 & 19.99 & 19.59 & 19.32 & 19.20 & 19.25 & 19.36 & 19.72 & 20.19 & 20.79 & 21.34 & 21.01 & 20.17 & 18.84 \\
 \nodata    & 0.256 & 0.197 & 0.206 & 0.187 & 0.250 & 0.258 & 0.433 & 0.724 & 1.455 & 2.639 & 2.292 & 1.480 & 0.515 \\
CXOM31 J004237.9+410526 & 20.92 & 20.08 & 20.01 & 19.49 & 18.92 & 18.95 & 18.14 & 17.83 & 17.33 & 17.21 & 16.99 & 16.68 & 16.65 \\
 \nodata    & 0.436 & 0.287 & 0.385 & 0.260 & 0.206 & 0.208 & 0.127 & 0.107 & 0.083 & 0.081 & 0.101 & 0.069 & 0.080 \\
CXOM31 J004215.5+412031 & 19.21 & 18.97 & 18.70 & 18.81 & 18.54 & 18.65 & 18.61 & 18.52 & 18.61 & 18.61 & 19.55 & 18.28 & 18.58 \\
 \nodata    & 0.138 & 0.147 & 0.148 & 0.183 & 0.197 & 0.228 & 0.253 & 0.265 & 0.359 & 0.422 & 1.277 & 0.413 & 0.626 \\
CXOM31 J004226.1+412551 & 19.90 & 20.03 & 19.84 & 20.00 & 19.82 & 20.14 & 19.51 & 19.58 & 19.40 & 20.03 & 19.81 & 21.25 & 19.28 \\
 \nodata    & 0.151 & 0.199 & 0.211 & 0.277 & 0.354 & 0.476 & 0.326 & 0.408 & 0.475 & 0.899 & 1.003 & 4.407 & 0.847 \\
CXOM31 J004255.3+412558 & 20.48 & 19.51 & 19.34 & 19.16 & 18.67 & 18.62 & 18.65 & 18.49 & 18.33 & 18.42 & 17.97 & 18.04 & 18.71 \\
 \nodata    & 0.485 & 0.280 & 0.335 & 0.306 & 0.284 & 0.281 & 0.336 & 0.322 & 0.313 & 0.386 & 0.305 & 0.340 & 0.646 \\
CXOM31 J004353.6+411654 & 18.51 & 18.43 & 18.23 & 18.24 & 18.12 & 18.08 & 18.15 & 18.05 & 18.08 & 18.03 & 17.85 & 17.80 & 17.87 \\
 \nodata    & 0.088 & 0.088 & 0.078 & 0.086 & 0.104 & 0.094 & 0.116 & 0.122 & 0.154 & 0.156 & 0.162 & 0.163 & 0.213 \\
CXOM31 J004303.1+411015 & 16.69 & 15.98 & 15.63 & 15.52 & 15.06 & 14.97 & 14.81 & 14.72 & 14.65 & 14.58 & 14.46 & 14.44 & 14.46 \\
 \nodata    & 0.010 & 0.007 & 0.007 & 0.007 & 0.007 & 0.007 & 0.007 & 0.008 & 0.009 & 0.010 & 0.012 & 0.011 & 0.014 \\
\enddata
\end{deluxetable}

%% file: table5.tex
\begin{deluxetable}{ccccccccccc}
\tablenum{5} \rotate \tablewidth{0pt}
\tabletypesize{\footnotesize}
\tablecaption{The Identification of 4 X-ray
GC Candidates in M31}
\tablehead{\colhead{Name}&\colhead{R.A.}&\colhead{Dec.}&\colhead{FWHM(pixel)}&\colhead{FWHM(pixel)}&\colhead{Known X-ray}&\colhead{$\chi^2_{min}$}&\colhead{Q}&\colhead{Type}&\colhead{Radial}\\
\colhead{}&\colhead{(J2000)}&\colhead{(J2000)}&\colhead{of
Sources}&\colhead{of
stars}&\colhead{GCs}&\colhead{}&\colhead{}&\colhead{}&\colhead{Offset(\arcsec)}}
\tablecolumns{10} \startdata
 CXOM31 J004215.0+412122 &00 42 15.03  &+41 21 22.0 & $2.4\pm0.04$  &$2.1\pm0.04$ &Bo86,MIT164  &  3.66 &0.988  & GC & 4.15\\
 CXOM31 J004222.6+412234 &00 42 22.62  &+41 22 34.9 & \nodata       &\nodata      &\nodata      &\nodata&\nodata&\nodata& 3.60\\
 CXOM31 J004237.9+410526 &00 42 37.94  &+41 05 26.1 & $2.6\pm0.06$  &$2.5\pm0.06$ &Bo146 &149.20&0.000&     & 2.84\\
 CXOM31 J004215.5+412031 &00 42 15.53  &+41 20 31.9 & $3.3\pm0.11$  &$2.1\pm0.11$ &Bo7   &  5.51 &0.938& GC & 2.41\\
 CXOM31 J004226.1+412551 &00 42 26.11  &+41 25 51.0 & $2.6\pm0.06$  &$2.1\pm0.06$ &Bo7   & 31.79 &0.001&    & 0.47\\
 CXOM31 J004255.3+412558 &00 42 55.31  &+41 25 58.0 & \nodata       &\nodata      &\nodata      &\nodata&\nodata&\nodata& 3.79\\
 CXOM31 J004353.6+411654 &00 43 53.62  &+41 16 54.0 & $4.5\pm0.10$  &$2.2\pm0.10$ &Bo7   & 10.83 &0.543& GC & 4.08\\
 CXOM31 J004303.1+411015 &00 43 03.08  &+41 10 16.1 & $2.4\pm0.08$  &$2.3\pm0.08$ &Bo138 &  4.84 &0.963& GC & 2.84\\
\enddata
\end{deluxetable}

%% file: table6.tex
\begin{deluxetable}{cccccc}
\tablenum{6} 
\tabletypesize{\footnotesize}
\tablewidth{0pt} \tablecaption{The Parameters of 62
X-ray GC Candidates}
\tablehead{\colhead{Source}&\colhead{Optical ID\tablenotemark{a}}&\colhead{X-ray ID\tablenotemark{b}}&\colhead{$L_{X}$\tablenotemark{c}}&\colhead{V\tablenotemark{d}}&\colhead{B-V\tablenotemark{e}}\\
\colhead{ID}&\colhead{}&\colhead{}&\colhead{$10^{35}\rm{erg~s^{-1}}$}&\colhead{}&\colhead{}}
\tablecolumns{6}
\startdata
  1\tablenotemark{1}  & G1              &                  & 6           & \nodata       & \nodata       \\
  2\tablenotemark{1}  & Bo293           &  RX J0036.3+4053 & 5.1         &16.39          & 0.45          \\
  3\tablenotemark{1}  & Bo5             &  SHP73           & 1990        &15.67          & 1.02          \\
  4\tablenotemark{1}  & MA94a(16)       &  RX J0040.5+4033 & 9.0         & \nodata       & \nodata       \\
  5                   & Bo7             &  RX J0040.4+4129 & 71.5        &14.25$\pm$0.003& 0.50$\pm$0.006\\
  6                   & MIT87           &  D27             & 3           &18.99$\pm$0.129& 1.02$\pm$0.220\\
  7                   & Bo45            &  RX J0041.7+4134 & 450         &15.74$\pm$0.008& 0.90$\pm$0.022\\
  8\tablenotemark{1}  & Bo58,MIT106     &  D22             & 1-24        &15.06          & 0.80          \\
  9                   & Bo55            &                  & 2.4-2.9     &16.67$\pm$0.031& 1.55$\pm$0.117\\
 10                   & Bo D42,MIT130   &  SHP138,D13      & 31-73       &18.62$\pm$0.092& 0.94$\pm$0.120\\
 11                   & Bo D44          &  D15             & 40-100      &18.45$\pm$0.085& 1.24$\pm$0.211\\
 12                   & MIT140          &  D10             & 40-114      &16.78$\pm$0.064& 1.64$\pm$0.158\\
 13                   & Bo78,MIT153     &  D20             & 24-300      &17.65$\pm$0.199& 1.71$\pm$0.378\\
 14                   & Bo82,MIT159     &  SHP150,D2       & 1735-2330   &15.48$\pm$0.009& 1.71$\pm$0.040\\
 15\tablenotemark{n}  &                 & CXOM31 J004215.0+412122   & 3           &18.31$\pm$0.085& 0.74$\pm$0.105\\
 16\tablenotemark{n}  &                 & CXOM31 J004215.5+412031   & 26          &18.64$\pm$0.219& 0.57$\pm$0.268\\
 17                   & Bo86,MIT164     &  SHP158,D4       & 498-763     &14.97$\pm$0.043& 0.72$\pm$0.058\\
 18                   & MIT165/MIT166   &  D24             & 5           &18.23$\pm$0.133& 1.66$\pm$0.440\\
 19                   & Bo94,MIT173     &  SHP168,D21      & 19          &15.52$\pm$0.010& 0.83$\pm$0.020\\
 20                   & Bo98            &  D25             & 1-12        &16.19$\pm$0.008& 0.84$\pm$0.011\\
 21                   & Bo96,MIT174     &  D14             & 28-174      &16.11$\pm$0.109& 1.20$\pm$0.166\\
 22                   & Bo D63          &  [WGK2004]s1-83  & 154         &18.52$\pm$0.104& 0.54$\pm$0.182\\
 23                   & Bo110           &  SHP178          & 41-52       &15.13$\pm$0.008& 0.81$\pm$0.015\\
 24                   & Bo117           &                  & 14          &16.22$\pm$0.032& 0.70$\pm$0.046\\
 25                   & Bo107,MIT192    &  SHP175,D16      & 56-290      &15.83$\pm$0.081& 0.93$\pm$0.114\\
 26                   & NBol 63         & CXOM31 J004231.2+412008   & 5          &16.73$\pm$0.170& 0.74$\pm$0.209\\
 27                   & Bo116           & RX J0042.5+4132  & 234         &16.79$\pm$0.023& 1.37$\pm$0.091\\
 28                   & Bo123,MIT212    &  D18             & 16-27       &17.28$\pm$0.199& 0.98$\pm$0.279\\
 29                   & MIT213          &  D23             & 34-137      &13.25$\pm$0.164& 1.06$\pm$0.225\\
 30                   & PB-in7          & CXOM31 J004246.0+411736   & 9           &14.66$\pm$0.273& 1.00$\pm$0.359\\
 31                   & Bo128           &  [PFJ93]51       & 237         &16.84$\pm$0.107& 0.85$\pm$0.142\\
 32\tablenotemark{2}  & MIT222          &  D28             & 3           & \nodata       & \nodata       \\
 33                   & Bo135           &  SHP205          & 3093-4009   &15.89$\pm$0.012& 0.92$\pm$0.027\\
 34                   & Bo138           &                  & 8-84        &15.84$\pm$0.206& 0.97$\pm$0.282\\
 35                   & Bo144           &  D6              & 216-512     &15.98$\pm$0.251& 1.00$\pm$0.329\\
 36                   & Bo143           &  SHP217,D5       & 152-555     &15.89$\pm$0.099& 0.85$\pm$0.128\\
 37\tablenotemark{n}  &                 & CXOM31 J004303.1+411015    & 62          &15.29$\pm$0.008& 1.05$\pm$0.011\\
 38                   & Bo146           &  SHP220,D7       & 74-414      &16.81$\pm$0.331& 0.81$\pm$0.426\\
 39                   & Bo D91,MIT236   &  SHP218          & 553-692     &14.26$\pm$0.005& 1.44$\pm$0.012\\
 40                   & Bo147,MIT240    &  SHP222,D11      & 52-194      &15.52$\pm$0.057& 0.90$\pm$0.082\\
 41                   & Bo148           &  SHP223,D8       & 105-418     &15.77$\pm$0.097& 0.78$\pm$0.123\\
 42                   & $\rm [WSB85]S3~14$    & CXOM31 J004304.2+411601   & 43          &16.61$\pm$0.180& 0.86$\pm$0.240\\
 43                   & Bo150,MIT246    &  D19             & 16-72       &16.24$\pm$0.097& 1.08$\pm$0.158\\
 44                   & $\rm [WSB85]S1~4$     & CXOM31 J004309.7+411901   & 212         &17.86$\pm$0.283& 1.55$\pm$0.530\\
 45                   & Bo153,MIT251    &  SHP228,D3       & 373-1248    &16.09$\pm$0.084& 1.02$\pm$0.120\\
 46                   & Bo158           &  SHP229,D12      & 600-1880    &14.60$\pm$0.004& 0.88$\pm$0.009\\
 47                   & Bo161,MIT260    &  D26             & 16-22       &16.23$\pm$0.035& 0.86$\pm$0.056\\
 48                   & Bo159           &                  & 2           &16.96$\pm$0.063& 1.20$\pm$0.128\\
 49                   & Bo164,MA94a(269)&  RX J0043.2+4112 & 13          &17.68$\pm$0.115& 0.95$\pm$0.186\\
 50                   & Bo163           &  RX J0043.2+4127 & 1-1010      &15.00$\pm$0.008& 0.98$\pm$0.016\\
 51                   & Bo182           &                  & 46          &15.35$\pm$0.007& 0.96$\pm$0.017\\
 52                   & Bo185,MIT299    &  SHP247,D9       & 454-1981    &15.49$\pm$0.011& 0.91$\pm$0.021\\
 53                   & MIT311          &  SHP250          & 52          &18.34$\pm$0.121& 1.14$\pm$0.350\\
 54                   & Bo193           &  SHP253          & 44          &15.33$\pm$0.008& 0.95$\pm$0.017\\
 55\tablenotemark{n}  &                 & CXOM31 J004353.6+411654   & 305         &18.22$\pm$0.112& 0.40$\pm$0.145\\
 56                   & Bo204           &  SHP261          & 37          &15.66$\pm$0.010& 0.86$\pm$0.022\\
 57                   & Bo213           &  2E 0041.3+4114           & 201         &16.98$\pm$0.024& 0.74$\pm$0.034\\
 58                   & MA94a(380)      &  RX J0044.4+4136 & 27          &14.72$\pm$0.005& 0.87$\pm$0.011\\
 59                   & Bo225           &  SHP282          & 1130        &14.13$\pm$0.003& 0.87$\pm$0.007\\
 60\tablenotemark{1}  & MA94a(447)      &  RX J0045.4+4132 & 15          & \nodata       & \nodata       \\
 61\tablenotemark{1}  & Bo375           &  SHP318,D1       & 5148-10372  &17.54          & 1.47          \\
 62\tablenotemark{1}  & Bo386           &  SHP349          & 1496        &15.59          & 0.76          \\
\enddata
\tablenotetext{a}{Source identifications beginning with Bo refer
to the GC candidates listed in \citet{batt87}; Identifications
beginning with Bo D refer to the GC candidates listed in
\citet{batt80}; MIT - in \citet{mag93}; MA94a - in \citet{ma94};
NBol - in \citet{batt93}; PB - in \citet{bp01}.}
\tablenotetext{b}{Source identifications beginning with SHP refer
to M31 $ROSAT$/PSPC X-ray source catalog entries in \citet{su97,
su01}; Identifications beginning with D refer to the list of GC
X-ray sources listed in \citet{dist02}; 2E - in \citet{bh00};
[PFJ93] - in \citet{pri93}; [WGK2004] - in \citet{wi04}; [WSB85] -
in \citet{wirth85}.} \tablenotetext{c}{No. 2, 4, 5, 7, 49, 58, 60
are in 0.1 - 2 keV; No. 15, 26, 30 are in 0.3 - 7 keV; No. 16, 22,
27, 37, 42, 44, 55 are in 0.1 - 10 keV; No. 31, 57 are in 0.2 -
4.0 keV. The others are in 0.3 - 10.0 keV.} \tablenotetext{d,e} {V
mags and B-V colors with error bars are from the BATC, and without
error bars are from \citet{batt87}.} \tablenotetext{n}{The new
X-ray GC candidates identified in this paper.}
\tablenotetext{1}{The sources are out of the BATC M31 field.}
\tablenotetext{2}{The sources are in the BATC M31 field but can
not be detected because of low signal-to-noise ratio.}

\end{deluxetable}